\documentclass[prd,twocolumn,aps,showpacs,preprintnumbers,amsmath,amssymb]{revtex4}
\usepackage{graphicx}
\usepackage{dcolumn}
\usepackage{bm}
\usepackage{psfrag}

\begin{document}
\newcommand{\be}{\begin{equation}}\newcommand{\ee}{\end{equation}}
\newcommand{\bea}{\begin{eqnarray}}\newcommand{\eea}{\end{eqnarray}}
\newcommand{\bc}{\begin{center}}\newcommand{\ec}{\end{center}}
\def\no{\nonumber}
\def\eq#1{Eq. (\ref{#1})}\def\eqeq#1#2{Eqs. (\ref{#1}) and  (\ref{#2})}
\def\lsim{\raise0.3ex\hbox{$\;<$\kern-0.75em\raise-1.1ex\hbox{$\sim\;$}}}
\def\gsim{\raise0.3ex\hbox{$\;>$\kern-0.75em\raise-1.1ex\hbox{$\sim\;$}}}
\def\slash#1{\ooalign{\hfil/\hfil\crcr$#1$}}
\def\eff{\mbox{\tiny{eff}}}
\def\order#1{{\mathcal{O}}(#1)}
\def\pppm{B^0\to\pi^+\pi^-}
\def\pzpz{B^0\to\pi^0\pi^0}
\def\pppz{B^0\to\pi^+\pi^0}
\preprint{ }
\title{$B\to K\pi$ decays}
\author{
T. N. Pham}
\affiliation{
 Centre de Physique Th\'{e}orique, CNRS \\ 
Ecole Polytechnique, 91128 Palaiseau, Cedex, France }
\date{\today}
\begin{abstract}
Though QCD Factorization(QCDF)
could produce  sufficiently large $B\to K\pi$ branching ratios close to
experiments, the predicted 
direct CP asymmetry in ${\bar B}^{0}\to K^{+}\pi^{-}$ decay
is however still quite below experiment, even with a large
negative phase in the annihilation term, according to existing 
calculations. This suggests the presence of an additional strong 
phase in the decay amplitude 
which could come from the long-distance final state interactions(FSI) 
like charming penguin. In this paper,
we show that, by adding to the $B\to K\pi$ decay QCDF amplitude,  a real part 
and   an  absorptive part with a strength $10\% $ and $30\%$  
of the penguin amplitude, respectively, we could bring the 
 $B\to K\pi$ branching ratios and the  ${\bar B}^{0}\to K^{-}\pi^{+}$  
CP asymmetry close to the measured values. We also find that 
 the color-allowed electroweak penguin  is appreciable and that from the QCDF
electroweak-strong penguin interference terms and the  measured 
$B^{-}\to {\bar K}^{0}\pi^{-}$  we obtain $(9.0\pm 0.3)\times 10^{-6}$
for the ${\bar B}^{0}\to {\bar K}^{0}\pi^{0}$ branching ratio, a bit
lower than experiment. Similarly, the predicted value
 for $B^{-}\to K^{-}\pi^{0}$ in terms of the ${\bar B}^{0}\to K^{-}\pi^{+}$
branching ratio agrees well with experiment. This suggests that the 
measured value for ${\bar B}^{0}\to {\bar K}^{0}\pi^{0}$
in fact should be lower than the current measured value.
\end{abstract}
\maketitle
\section{INTRODUCTION}
Existing calculations of the penguin-dominated $B\to K\pi$ decays  
in QCD Factorization(QCDF) \cite{QCDF1,QCDF2,Zhu1} seem to produce 
$B\to K\pi$ branching ratios  mre or less in agreement with experiments, 
with a moderate annihilation contribution, 
but the predicted direct CP asymmetry in ${\bar B}^{0}\to K^{-}\pi^{+}$
decay is smaller than experiment by a factor of 2
or more, even with a large negative strong phase for the annihilation
terms as in scenario S4 \cite{QCDF2}. In fact in QCDF, and
also in naive Factorization model \cite{Ali}, the absorptive part(the
imaginary part) of the short-distance Wilson coefficients in the 
non leptonic decay effective Hamiltonian  are negative and of the same sign
as the real part, the direct CP asymmetry for
${\bar B}^{0}\to \pi^{+}\pi^{-}$ and ${\bar B}^{0}\to K^{-}\pi^{+}$ are
negative and positive, respectively, in opposite sign with 
measurements\cite{Cheng1}.
To reverse the sign of the CP asymmetry and to
produce a large CP asymmetry, one would need a large annihilation
term with a large negative phase. Another possibility is   
a long-distance final state interaction(FSI) term, probably mainly 
absorptive, to generate a large strong phase, but 
not to overestimate the  branching ratios, like  charming penguin 
due to $B \to D_{s} D\to  K\pi$ and $B \to D_{s}^{*} D^{*}\to  K\pi$
inelastic FSI effect \cite{Isola,Pham1,Ciuchini,Fazio,Fazio2,Cheng} 
which are CKM-favored, color-allowed processes and could produce
a large absorptive part  in $B\to K\pi$ decays.
In this paper, we show that
by adding to the QCDF amplitude, a penguin-like amplitude with a
real and absorptive part with a strength $10\% $ and $30\%$, respectively, 
of the penguin amplitude, we could obtain the $B\to K\pi$
branching ratios and the  ${\bar B}^{0}\to K^{-}\pi^{+}$  
CP asymmetry close to experiments.
We also find that the color-allowed electroweak penguin(EW) from 
the operators  $O_{7}$ and $O_{9}$ contribute
appreciably through the  strong penguin-EW penguin 
interference terms in $B^{-}\to K^{-}\pi^{0}$ and 
${\bar B}^{0}\to {\bar K}^{0}\pi^{0}$ decays. Then
using only the QCDF expressions for the color-allowed tree-penguin(T-P),  
color-suppressed tree-penguin(C-P), color-allowed electroweak-penguin(EW-P)
interference terms and the measured 
${\mathcal B}(B^{-}\to {\bar K}^{0}\pi^{-})$ and
${\mathcal B}({\bar B}^{0}\to K^{-}\pi^{+})$, we could predict    
${\mathcal B}(B^{-}\to  K^{-}\pi^{0})$ and ${\mathcal B}({\bar B}^{0}\to {\bar K}^{0}\pi^{0})$. We find that the predicted 
${\mathcal B}( B^{-}\to K^{-}\pi^{0})$  agrees well with experiment 
while the predicted ${\mathcal B}({\bar B}^{0}\to K^{0}\pi^{0})$ is 
below the measured value. This strongly suggests that the measured 
${\mathcal B}({\bar B}^{0}\to {\bar  K}^{0}\pi^{0})$ is a bit 
too high as shown  in the following. In the next section we will  
present numerical results for the 
$B\to K\pi$ branching ratios and CP asymmetry in QCDF factorization
with the addition of  a small penguin-like amplitude to generate a  CP
asymmetry for ${\bar B}^{0}\to K^{-}\pi^{+}$ decay. We then give prediction
for ${\mathcal B}( B^{-}\to K^{-}\pi^{0})$ and ${\mathcal B}({\bar
  B}^{0}\to K^{0}\pi^{0})$ from the electroweak penguin and strong 
penguin interference terms.

\section{$B\to K\pi$   DECAY IN QCD FACTORIZATION}
The $B\to M_{1} M_{2}$ decay amplitude in QCDF is given by \cite{QCDF1,QCDF2}: 
\bea
&&  {\cal A}(B \rightarrow M_1 M_2)=
 \frac{G_F}{\sqrt{2}}\sum_{p=u,c}V_{pb}V^{*}_{ps}\times  \nonumber \\
&& \left( -\sum_{i=1}^{10} a_i^p
   \langle M_1 M_2 \vert O_i \vert B \rangle_H + 
 \sum_{i}^{10} f_B f_{M_1}f_{M_2} b_i \right ),
\label{BMM}
\eea 
where the QCD coefficients  $a_{i}^{p}$ contain the vertex corrections,
penguin corrections, and hard spectator scattering contributions, 
the hadronic matrix elements $ \langle M_1 M_2 \vert O_i \vert B
\rangle_H $  of the tree and penguin operators $O_{i}$ are given 
by factorization model \cite{Ali,Zhu2}, $b_{i}$ are annihilation
 contributions. The values for $a_{i}^{p}$,$p=u,c$ , computed from 
the expressions in \cite{QCDF1,QCDF2} at the renormalization 
scale $\mu=m_{b}$, with $m_{b}=4.2\,\rm GeV$, are~:
\bea
&& a_{4}^{c}=-0.033 - 0.013\,i + 0.0009\,\rho_{H}\exp(i\phi_{H}),\nonumber \\
&& a_{4}^{u}=-0.027 - 0.017\,i + 0.0009\,\rho_{H}\exp(i\phi_{H}),\nonumber \\
&& a_{6}^{c}=-0.045 - 0.003\,i ,\quad a_{6}^{u}=-0.042 - 0.013\,i ,\nonumber \\
&& a_{8}^{c}=-0.0004 - 0.0001\,i ,\quad  a_{8}^{u}= 0.0004 - 0.0001\,i ,\nonumber \\
&& a_{10}^{c}=-0.0011 - 0.0001\,i - 0.0006\,\rho_{H}\exp(i\phi_{H}) ,\nonumber \\
&& a_{10}^{u}=-0.0011 + 0.0006\,i - 0.0006\,\rho_{H}\exp(i\phi_{H}).
\label{aiuc}
\eea
for $i=4,6,8,10$. For other coefficients, $a_{i}^{u}=a_{i}^{p}=a_{i}$ :
\bea
&& a_{1}= 1.02 + 0.015\,i -0.012\,\rho_{H}\exp(i\phi_{H}),\nonumber \\
&& a_{2}= 0.156 - 0.089\,i + 0.074\,\rho_{H}\exp(i\phi_{H}), \nonumber \\
&& a_{3}= 0.0025 + 0.0030\,i - 0.0024\,\rho_{H}\exp(i\phi_{H}), \nonumber \\
&& a_{5}=-0.0016 - 0.0034\,i + 0.0029\,\rho_{H}\exp(i\phi_{H}), \nonumber \\
&& a_{7} =-0.00003 - 0.00004\,i - 0.00003\,\rho_{H}\exp(i\phi_{H})\nonumber \\
&& a_{9} = -0.009 - 0.0001\,i + 0.0001\,\rho_{H}\exp(i\phi_{H}).
\label{ai}
\eea
where the complex parameter $\rho_{H}\exp(i\phi_{H})$ represents the 
end-point singularity contribution in the hard-scattering
corrections $X_{H}=(1
+\rho_{H}\exp(i\phi_{H}))\,\ln(\frac{m_{B}}{\Lambda_{h}})$ \cite{QCDF1,QCDF2}. 

For the annihilation terms, we have~:
\bea
 b_{2}\kern -0.2cm&=&\kern -0.2cm -0.0041 -\kern -0.1cm 0.0071\rho_{A}\exp(i\phi_{A}) - 0.0019(\rho_{A}\exp(i\phi_{A}))^{2} ,\nonumber \\
 b_{3}\kern -0.2cm&=& \kern -0.2cm -0.0071 -\kern -0.1cm 0.016\rho_{A}\exp(i\phi_{A}) - 0.0093(\rho_{A}\exp(i\phi_{A}))^{2}, \nonumber \\
 b_{3}^{ew}\kern -0.2cm&=&\kern -0.2cm -0.00012  -0.00016\,
\rho_{A}\exp(i\phi_{A})\nonumber \\
 &+&\kern -0.2cm 0.000003\,(\rho_{A}\exp(i\phi_{A}))^{2}.
\label{b3}
\eea
where $b_{i}$ are evaluated with the factor $f_{B}f_{M_{1}}f_{M_{2}}$
included and normalized relative to the factor
$f_{K}F^{B\pi}_{0}(m_{B}^{2}-m_{\pi}^{2})$ in the factorisable terms,
and $\rho_{A}$ , like $\rho_{H}$, appears in the divergent 
annihilation term 
$X_{A}=(1 +\rho_{A}\exp(i\phi_{A}))\,\ln(\frac{m_{B}}{\Lambda_{h}})$.

The  $B\to K\pi$ decay amplitude with the factorisable part \cite{Ali}
and the annihilation term \cite{QCDF1,QCDF2,Zhu3} are:
\bea
\kern -0.6cm&&A(B^{-}\to K^{-}\pi^{0}) = -i\frac{G_{F}}{2}f_{K}F^{B\pi}_{0}(m^{2}_{K})
(m_{B}^{2}-m_{\pi}^{2})\nonumber \\
\kern -0.6cm&&\left(V_{ub}V^{*}_{us}a_{1}
+(V_{ub}V^{*}_{us}+V_{cb}V^{*}_{cs})[a_{4} + a_{10} + (a_{6}+a_{8})r_{\chi}]\right)\nonumber \\ 
\kern -0.6cm&& -i\frac{G_{F}}{2}f_{\pi}F^{BK}_{0}(m^{2}_{\pi})(m_{B}^{2}-m_{K}^{2})\nonumber \\ 
\kern -0.6cm
&&\times\left(V_{ub}V^{*}_{us}a_{2}+(V_{ub}V^{*}_{us}+V_{cb}V^{*}_{cs})\times
  \frac{3}{2}(a_{9}-a_{7})\right)\nonumber \\ 
\kern -0.6cm&&-i\frac{G_{F}}{2}f_{B}f_{K}f_{\pi}\nonumber \\ 
\kern -0.6cm&&\times\left[V_{ub}V^{*}_{us}b_{2} 
+ (V_{ub}V^{*}_{us}+V_{cb}V^{*}_{cs})\times(b_{3} + b_{3}^{ew})\right]
\label{K1p0} \\
\kern -0.6cm&&A(B^{-}\to {\bar K}^{0}\pi^{-}) =-i\frac{G_{F}}{\sqrt{2}}f_{K}F^{B\pi}_{0}(m^{2}_{K})(m_{B}^{2}-m_{\pi}^{2})\nonumber \\
\kern -0.6cm&&+(V_{ub}V^{*}_{us}+V_{cb}V^{*}_{cs})\left[a_{4} - \frac{1}{2}a_{10} + (a_{6}-\frac{1}{2}a_{8})r_{\chi}\right]\nonumber \\
\kern -0.6cm&&-i\frac{G_{F}}{\sqrt{2}}f_{B}f_{K}f_{\pi}\nonumber \\ 
\kern -0.6cm&&\times\left[V_{ub}V^{*}_{us}b_{2} 
+ (V_{ub}V^{*}_{us}+V_{cb}V^{*}_{cs})\times(b_{3} + b_{3}^{ew})\right]
\label{K0p1}
\eea
and for ${\bar B}^{0}$ :
\bea
\kern -0.6cm&&A({\bar B}^{0}\to K^{-}\pi^{+}) = -i\frac{G_{F}}{\sqrt{2}}f_{K}F^{B\pi}_{0}(m^{2}_{K})(m_{B}^{2}-m_{\pi}^{2})\nonumber \\
\kern -0.6cm&&\biggl(\kern -0.1cm V_{ub}V^{*}_{us}a_{1}+\kern -0.1cm (V_{ub}V^{*}_{us}+\kern -0.1cm
  V_{cb}V^{*}_{cs})[a_{4} +\kern -0.1cm a_{10} + \kern
-0.1cm(a_{6}+a_{8})r_{\chi}]\kern -0.1cm\biggr)\nonumber\\
\kern -0.6cm&&-i\frac{G_{F}}{\sqrt{2}}f_{B}f_{K}f_{\pi}\left[
 (V_{ub}V^{*}_{us}+V_{cb}V^{*}_{cs})\times(b_{3} - \frac{b_{3}^{ew}}{2})\right]
\label{K1p2}\\
\kern -0.6cm&&A({\bar B}^{0}\to {\bar K}^{0}\pi^{0}) =i\frac{G_{F}}{2}f_{K}F^{B\pi}_{0}(m^{2}_{K})(m_{B}^{2}-m_{\pi}^{2})\nonumber \\
\kern -0.6cm&&\times(V_{ub}V^{*}_{us}+V_{cb}V^{*}_{cs})\left[a_{4} - \frac{1}{2}a_{10} + (a_{6}-\frac{1}{2}a_{8})r_{\chi}\right]\nonumber \\
\kern -0.6cm&& -i\frac{G_{F}}{2}f_{\pi}F^{BK}_{0}(m^{2}_{\pi})(m_{B}^{2}-m_{K}^{2})\nonumber \\ 
\kern -0.6cm&&\left(V_{ub}V^{*}_{us}a_{2}+(V_{ub}V^{*}_{us}+V_{cb}V^{*}_{cs})\times \frac{3}{2}(a_{9}-a_{7})\right)\nonumber \\ 
\kern -0.6cm&&+i\frac{G_{F}}{2}f_{B}f_{K}f_{\pi}\left[
 (V_{ub}V^{*}_{us}+V_{cb}V^{*}_{cs})\times(b_{3} - \frac{b_{3}^{ew}}{2})\right]
\label{K0p0}
\eea
where $r_{\chi}=\frac{2m_{K}^{2}}{(m_{b}-m_{d})(m_{d} + m_{s})}$
is the chirally-enhanced terms in the penguin $O_{6}$ matrix
element. 
We also need the $B^{-}\to \pi^{-}\pi^{0} $ amplitude:
\bea
\kern -0.6cm &&A(B^{-}\to \pi^{-}\pi^{0}) = -i\frac{G_{F}}{2}f_{\pi}F^{B\pi}_{0}(m^{2}_{\pi})
(m_{B}^{2}-m_{\pi}^{2})\nonumber \\
&&\biggl(V_{ub}V^{*}_{ud}(a_{1} + a_{2})
+(V_{ub}V^{*}_{ud} +V_{cb}V^{*}_{cd})\nonumber \\
\kern -0.6cm&&\times \frac{3}{2}(a_{9}-a_{7}+ a_{10} +a_{8}r_{\chi})\biggr)
\label{p2p1}
\eea

We see that the $B\to K\pi$ decay
amplitudes consist of a QCD penguin(P) $a_{4} + a_{6}r_{\chi} $ ,
a color-allowed tree(T) $a_{1}$, a color-suppressed  tree(C) $a_{2}$
, a color-allowed electroweak penguin(EW) $a_{9}-a_{7}$, a 
color-suppressed  electroweak penguin(EWC) $a_{10}+ a_{8}r_{\chi}$ terms. 
Because of the relative large Wilson coefficients, the QCD penguin, the  
color-allowed tree and the color-allowed electroweak contributions 
give the major contribution in $B\to K\pi$ decays.

For the CKM matrix elements, since the inclusive and exclusive data on
$|V_{ub}|$ differ by a large amount and  the higher inclusive  data
exceeds the unitarity limit for 
$R_{b} = |V_{ud}V_{ub}^{*}|/|V_{cd}V_{cb}^{*}| $ with the current value
$\sin(2\beta)=0.681\pm 0.025 $ \cite{PDG}, we shall determine $|V_{ub}|$ from
the more precise $|V_{cb}|$ data. We have \cite{CKM}:
\be
\kern -0.5cm \vert V_{ub}\vert=  \frac{\vert V_{cb}V_{cd}^{*}\vert}{\vert V_{ud}^{*}\vert} \vert  \sin \beta 
\sqrt{1+\frac{\cos^2 \alpha}{\sin^2 \alpha}} .
\label{Vub}
\ee
 With $\alpha=(99^{+13}_{-9})^{\circ}$ \cite{PDG} and 
$\vert V_{cb}\vert = (41.78\pm 0.30\pm 0.08)\times 10^{-3}$ \cite{Barberio}, we find
\be
\vert V_{ub}\vert = 3.60\times 10^{-3}.
\label{Vub1}
\ee
in good agreement with the exclusive data in the range
$\vert V_{ub}\vert = 3.33 -3.51$ \cite{Barberio} . A recent UT fit also gives 
$\vert V_{ub}\vert = (3.60\pm 0.12)\times 10^{-3} $ \cite{Bona}.
The measurements of the $B_{s}-\bar{B}_{s}$ mixing 
also allow the extraction of $|V_{td}|$ from $B_{d}-\bar{B}_{d}$
mixing data. The  current determination \cite{Abulencia} gives
$|V_{td}/V_{ts}|= (0.208^{+0.008}_{-0.006})$ which in turn can be used to
determined the  angle $\gamma$ from the unitarity relation \cite{CKM}:
\be
\kern -0.5cm \vert V_{td}\vert= \frac{\vert V_{cb}V_{cd}^{*}\vert}{\vert V_{tb}^{*}\vert} \vert  \sin \gamma
\sqrt{1+\frac{\cos^2 \alpha}{\sin^2 \alpha}}.
\label{Vtd}
\ee
with $|V_{tb}|=1 $, we find $\gamma = 66^{\circ} $
which implies an angle $\alpha = 91.8^{\circ}$, in good agreement 
with the value found in the current UT-fit value of $(88\pm 16)^{\circ}$
\cite{Brown}. In the following in our $B\to K\pi$ decay calculations, we shall
use the unitarity triangle values for $\vert V_{ub}\vert $ and $\gamma$. For 
other hadronic parameters we use the values in 
Table 1  of \cite{QCDF2} and take $m_{s}(\rm 2\,GeV)=80\,\rm MeV$.
For the $B\to \pi$ and $B\to K$ transition form factor, we use the current
light-cone sum rules central value \cite{Zwicky}~:
\be
F^{B\pi}_{0}(0)= 0.258,\quad   F^{BK}_{0}(0) = 0.33
\label{FBK}
\ee

\begin{figure*}[t]
\begin{center}
\includegraphics[width=7cm]{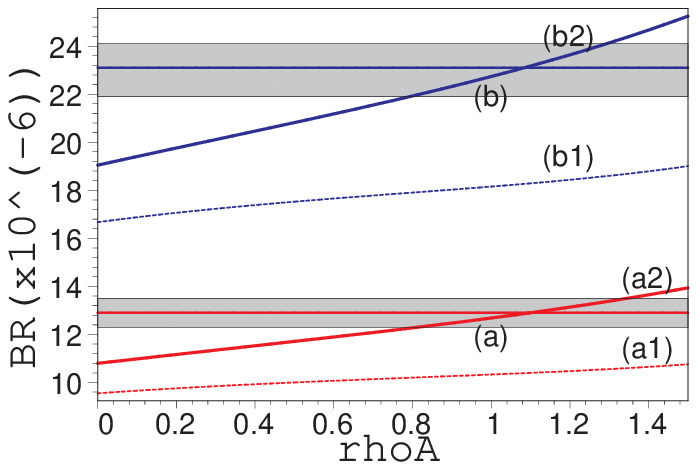}\hspace*{1cm}
\includegraphics[width=7cm]{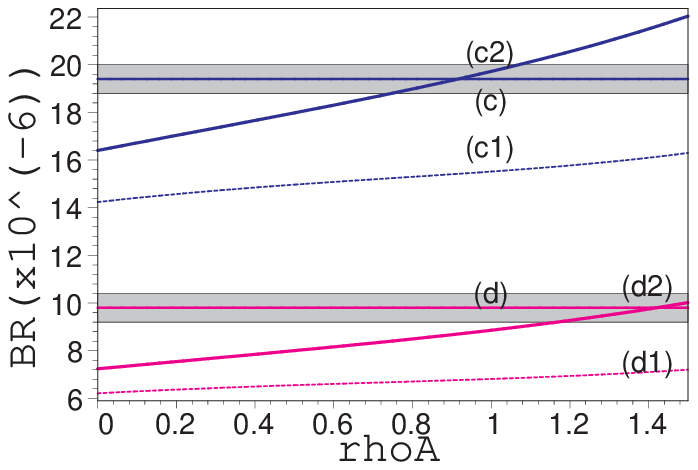} \vspace*{.5cm}
\caption{The computed and measured CP average branching ratios. The
horizontal line are the measured values \cite{HFAG} with the
gray areas represent the experimental errors. (a), (b), (c), (d) in 
the left and right figure represent the values for $B^{-}\to K^{-}\pi^{0}$, 
$B^{-}\to {\bar K}^{0}\pi^{-}$, ${\bar B}^{0}\to K^{-}\pi^{+}$ and 
${\bar B}^{0}\to {\bar K}^{0}\pi^{0}$ respectively. The curves (a1)-(d1) and
(a2)-(d2) are the corresponding QCDF predicted values for
$\phi_{A}=-55^{\circ}$, without and with additional penguin 
contribution respectively.
}
\label{fig:1}
\end{center}
\end{figure*}

\begin{figure*}[t]
\begin{center}
\includegraphics[width=7cm]{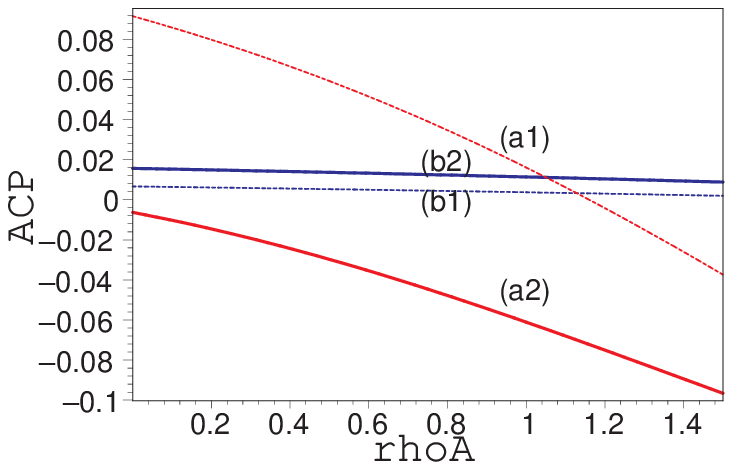}\hspace*{1cm}
\includegraphics[width=7cm]{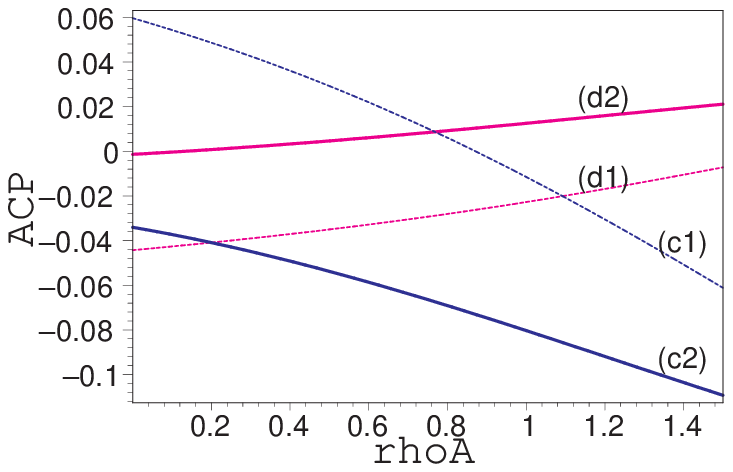}\vspace*{.5cm}
\caption{The same as in Fig.\ref{fig:1} but for the computed CP asymmetries.}
\label{fig:2}
\end{center}
\end{figure*}

  The computed branching ratios and direct CP asymmetries, with
$\rho_{H}=1,\phi_{H}=0$ and $\phi_{A}=-55^{\circ}$ as in scenario S4 
of \cite{QCDF2} are shown in Fig.\ref{fig:1} and Fig.\ref{fig:2} 
as function of $\rho_{A}$. For convenience we  also give in 
Table \ref{tab-res1} and Table \ref{tab-res2} the computed values 
at $\rho_{A}=1$ as in S4 with and without the additional 
penguin-like contribution $\delta P$.

\begin{center}
\begin{table}[h]
\begin{tabular}{|c|c|c|c|}

\hline
Modes & $\delta P=0$&$\delta P\not \!=0$   & Exp \cite{HFAG} \\
\hline
$B^{-}\to \pi^{-}\pi^{0}$ &$5.7$ &$5.7$ & $5.59\pm 0.4$\\
$B^{-}\to K^{-}\pi^{0}$ &$10.3$ &$12.7$ & $12.9\pm 0.6$\\
$B^{-}\to {\bar K}^{0}\pi^{-}$&$18.1$ &$22.7$ & $23.1\pm 1.0$\\
$\bar{B}^{0}\to  K^{-}\pi^{+}$ &$15.5$ &$19.7$ & $19.4 \pm 0.6$\\
${\bar B}^{0}\to {\bar K}^{0}\pi^{0}$&$6.8$ &$8.8$ & $9.8\pm 0.6$\\
\hline

\end{tabular}
\caption{ The CP-averaged $B\to K\pi$
Branching ratios in in unit of $ 10^{-6}$  in QCDF with and without
additional penguin-like contribution $\delta P$ and with
 $\rho_{A}=1.0$, $\phi_{A}= -55^{\circ}$}\label{tab-res1}
\end{table}
\end{center}

\begin{center}
\begin{table}[h]
\begin{tabular}{|c|c|c|c|}

\hline
Modes & $\delta P=0$&$\delta P\not \!=0$   & Exp \cite{HFAG} \\
\hline
$B^{-}\to \pi^{-}\pi^{0}$ &$0.0$ &$0.0$ & $0.06\pm 0.05$\\
$B^{-}\to K^{-}\pi^{0}$ &$0.02$ &$-0.06$ & $0.05\pm 0.025$\\
$B^{-}\to {\bar K}^{0}\pi^{-}$&$0.004$ &$0.01$ & $-0.009\pm 0.025$\\
$\bar{B}^{0}\to  K^{-}\pi^{+}$ &$-0.01$ &$-0.08$ & $-0.098^{+0.012}_{-0.010}$\\
${\bar B}^{0}\to {\bar K}^{0}\pi^{0}$&$-0.02$ &$0.01$ & $-0.01\pm 0.0.10$\\
\hline

\end{tabular}
\caption{ The direct $B\to K\pi$ CP asymmetries  in QCDF with and without
additional penguin-like contribution $\delta P$ and with
 $\rho_{A}=1.0$, $\phi_{A}= -55^{\circ}$}\label{tab-res2}
\end{table}
\end{center}
As can be seen, without the additional penguin contribution, the
computed branching ratios are slightly below experiments, but 
the CP asymmetries for ${\bar B}^{0}\to K^{-}\pi^{+}$
is much smaller than the measured values. Considering various theoretical
uncertainties, one could say that the real part of the $B\to K\pi$
decay amplitudes is more or less well described by QCDF, but the absorptive
part of the amplitude needs some additional contribution to account for
the observed CP asymmetries. As mentioned earlier, the factorisable 
contribution in the QCDF amplitude give the wrong sign for the CP 
asymmetries in ${\bar B}^{0}\to K^{-}\pi^{+}$, one needs a sizeable 
absorptive part to reverse the sign of the asymmetries. As shown in   
Fig.\ref{fig:1} and Fig.\ref{fig:2}, we would need an annihilation 
term with a very large $\rho_{A}$ to  produce correct $B\to K\pi$ 
branching ratios and  ${\bar B}^{0}\to K^{-}\pi^{+}$ asymmetry. 
This leaves us with charming penguin, as mentioned earlier, as a 
possible contribution to the large strong phase in $B\to K\pi$ 
decays. This possibility is considered in this paper by adding 
to the QCD penguin a small term: 
\be
\delta P = (ra_{4} + ra_{6}r_{\chi})(d_{1} +id_{2})
\label{dP}
\ee
where $ra_{4} $ and $ra_{6} $ are  the real part of 
$a_{4}$ and $a_{6}$, respectively. We note a previous fit with charming
penguin for $B\to PP, PV$  in \cite{Cottingham}. The results  shown 
above are obtained
with $d_{1}=0.08\ , d_{2}=-1/3$. We see that the branching ratios for 
${\bar B}^{0}\to K^{-}\pi^{+}$, $ B^{-}\to K^{-}\pi^{0}$ and 
$B^{-}\to {\bar K}^{0}\pi^{-}$ and the  CP asymmetry for 
${\bar B}^{0}\to K^{-}\pi^{+}$ are in good agreement with experiment. The
${\mathcal B}({\bar B}^{0}\to {\bar K}^{0}\pi^{0})$ is slightly below the measured 
value. Our results seem reasonable, as the predicted 
${\mathcal B}(B^{-}\to \pi^{-}\pi^{0}) $ which is sensitive $F^{B\pi}_{0}(0) $
agrees well with the measured value.  

 Thus QCDF with mainly absorptive additional penguin contribution seems to
describe rather well the $B\to K\pi$ observed branching ratios and the  
${\bar B}^{0}\to K^{-}\pi^{+}$ direct CP asymmetry. As all the 
$B\to K\pi$ branching ratios have been measured with an accuracy at the level
of $O(10^{-6})$, one could look at the difference in branching ratios
to test QCDF predictions for the electroweak and tree-level 
contributions. The color-favored electroweak penguin is CKM-favored, so 
its interference with the strong-penguin terms is about the same size 
as that of the color-favored tree and penguin interference. For example, from
Eq.(\ref{K1p0}-\ref{K0p0}), one sees that 
the differences  $2\Gamma( B^{-}\to K^{-}\pi^{0})-\Gamma({\bar B}^{0}\to
K^{-}\pi^{+})$ and 
$2\Gamma({\bar B}^{0}\to {\bar K}^{0}\pi^{0})-\Gamma(B^{-}\to {\bar K}^{0}\pi^{-})$ are
essentially given by the color-allowed electroweak and strong penguin 
interference term, so with QCDF expression for the interference 
term, one could predict ${\mathcal B}({\bar B}^{0}\to
{\bar K}^{0}\pi^{0})$ in terms of ${\mathcal B}(B^{-}\to {\bar K}^{0}\pi^{-}) $
or vice versa. We have:  
\bea
\kern -0.6 cm&&{\mathcal B}({\bar B}^{0}\to {\bar K}^{0}\pi^{0})=
[{\mathcal B}(B^{-}\to {\bar K}^{0}\pi^{-}) r_{b} + \delta B_{1}]/2
\label{BK0p0}\\
\kern -0.6 cm&&{\mathcal B}(B^{-}\to   K^{-}\pi^{0})=[{\mathcal B}({\bar B}^{0}\to
K^{-}\pi^{+})+ \delta B_{2}]/2r_{b}\\
\label{BK1p0}
\kern -0.6 cm&&{\mathcal B}({\bar B}^{0}\to  K^{-}\pi^{+})=[{\mathcal B}(B^{-}\to {\bar K}^{0}\pi^{-}) r_{b} + \delta B_{3}]
\label{Bk2p1}
\eea
and 
where $\delta B_{1}$, $\delta B_{2}$, $\delta B_{3}$ are respectively the
computed difference $2{\mathcal B}({\bar B}^{0}\to {\bar K}^{0}\pi^{0})-
r_{b}{\mathcal B}(B^{-}\to {\bar K}^{0}\pi^{-}) $, $2\,r_{b}{\mathcal B}(B^{-}\to
K^{-}\pi^{0})- {\mathcal B}({\bar B}^{0}\to  K^{-}\pi^{+})$, 
${\mathcal B}({\bar B}^{0}\to K^{-}\pi^{+} )-
r_{b}{\mathcal B}(B^{-}\to {\bar K}^{0}\pi^{-}) $ and
$r_{b}=\tau_{B^0}/\tau_{B^-}$. We find, in unit of $10^{-6}$, with 
$\rho_{A}=1, \phi_{A}=-55^{\circ}$ as with the computed values in Table
\ref{tab-res1} and Table \ref{tab-res2} (only  leading terms are shown):
\bea
\kern -0.6 cm&& \delta B_{1}= -3.52\,\left[-3.84({\rm EW}) + 1.42\cos(\gamma)({\rm C})\right]\label{dB1} \\
\kern -0.6 cm&& \delta B_{2}= 3.97\,\left[3.84({\rm EW}) -
  1.42\cos(\gamma)({\rm C})\right]
\label{dB2}\\
\kern -0.6 cm&&\delta B_{3}= -1.52\,[ - 5.53\cos(\gamma)({\rm TP})\nonumber\\
\kern -0.6 cm&& -0.65\cos(\gamma)({\rm TA}) + 0.69({\rm EWC}) + 0.36({\rm
  TT}) ]
\label{dB3}
\eea
In the above expressions, the main contributions are from the interference
 of the  color-allowed tree, color-suppressed tree, 
and electroweak penguin  with the strong penguin terms
and are shown inside the brackets in the above expressions. Then from
Eq.(\ref{BK0p0}-\ref{BK1p0}), we obtain (in unit of $10^{-6}$)
\bea
\kern -0.6 cm&&{\mathcal B}({\bar B}^{0}\to {\bar K}^{0}\pi^{0})= 9.0\pm 0.3, 
\label{BK0}\\
\kern -0.6 cm&&{\mathcal B}(B^{-}\to   K^{-}\pi^{0})= 12.5\pm 0.3.
\label{BK1}\\
\kern -0.6 cm&&{\mathcal B}({\bar B}^{0}\to  K^{-}\pi^{+})= 20.0\pm 0.6, 
\label{BK3}
\eea

which agrees with experiment, to within the current accuracy. In particular
the agreement with experiment for ${\mathcal B}(B^{-}\to
K^{-}\pi^{0}) $ shows that  the electroweak-penguin interference 
observed in $B\to K\pi$ decay rates difference is well described by  
QCDF. The predicted 
${\mathcal B}({\bar B}^{0}\to {\bar K}^{0}\pi^{0}) $ is a bit lower than
experiment, but this prediction is on firm ground since 
$dB_{1}\approx -dB_{2}$ and the predicted
${\mathcal B}(B^{-}\to   K^{-}\pi^{0}) $ in terms of $dB_{2}$ agrees well
with experiment, as mentioned above. 
In fact the discrepancy between experiment and the above 
prediction for ${\mathcal B}({\bar B}^{0}\to {\bar K}^{0}\pi^{0}) $
could be understood in the following way~:  First of all, 
Eqs.(\ref{dB1}-\ref{dB3})
shows that the electroweak-penguin as well as the color-suppressed 
tree with the strong penguin interference terms in 
${\mathcal B}(B^{-}\to  K^{-}\pi^{0}) $ and in ${\mathcal B}({\bar
  B}^{0}\to {\bar K}^{0}\pi^{0}) $ are essentially  the same in magnitude but
opposite in sign. This can be seen
from the decay amplitudes given in  Eqs.(\ref{K1p0} -\ref{K0p0}). Then 
these interference terms cancel out  leaving only the color-allowed
tree-penguin interference term in the sum
of  $\Gamma(B^{-}\to  K^{-}\pi^{0}) $ and $\Gamma({\bar B}^{0}\to {\bar
  K}^{0}\pi^{0}) $. Since the color-allowed tree-penguin interference
terms, the penguin and annihilation terms are essentially the same in
the sum  $ 2( \Gamma(B^{-}\to  K^{-}\pi^{0}) + \Gamma({\bar
  B}^{0}\to {\bar K}^{0}\pi^{0}) $ and ($\Gamma(B^{-}\to {\bar
  K}^{0}\pi^{-}+ \Gamma({\bar B}^{0}\to  K^{-}\pi^{+})$), by 
neglecting small quadratic term in tree and electroweak
penguin contributions, we obtain the approximate relation:
\bea
&&2[ {\mathcal B}(B^{-}\to  K^{-}\pi^{0})r_{b} + {\mathcal B}({\bar
  B}^{0}\to {\bar K}^{0}\pi^{0})]\nonumber\\
&& = [{\mathcal B}(B^{-}\to {\bar
  K}^{0}\pi^{-})r_{b}+ {\mathcal B}({\bar B}^{0}\to  K^{-}\pi^{+})]
\label{sum}
\eea
This relation has been first given 
in \cite{Lipkin,Rosner}, but was derived independently
in \cite{Isola1} from the isospin  $B\to K\pi$ amplitudes:
\bea
\Delta & = &\left\{ \Gamma(B^{-} \to \bar{K}^{0} {\pi}^{-})+\Gamma(\bar{B}^{0}
\to K^- {\pi}^+) \right. \nonumber\\
   & - & \left. 2\left[\Gamma(B^{-} \to K^- {\pi}^0)+\Gamma(\bar{B}^{0} \to
     \bar{K}^{0} {\pi}^{0})\right]\right\}\tau_{B^0}\nonumber\\
& = & \left[-{4\over 3}|B_3|^2-{8\over {\sqrt 3}}{\rm Re} (B_3^* B_1 {\rm
  e}^{i\delta})\right](C\tau_{{\bar B}^0})\,\, 
\label{delta}
\eea
where $B_{1}, B_{3}$ are respectively the isospin $I=1/2$ and $I=3/2$
tree-level $B\to K\pi$ amplitudes.
With the measured branching ratios, we get $(43.69\pm 2.4)\times 10^{-6}
$ and $(40.9\pm 2.0) \times 10^{-6}$ for the l.h.s and r.h.s of
Eq.(\ref{sum}), and a difference  
$2.72\times 10^{-6} $ between the two values. Ignoring the errors, we see that 
this difference is rather large, of the size of the tree-penguin interference 
term, while, theoretically, it should be of  $O(10^{-6}) $ or less and
has a value given by $\delta B_{1}+ \delta B_{2}=0.43\times 10^{-6}$.  
The  problem, we suspect, seems to be the measured  
${\mathcal B}( {\bar B}^{0}\to {\bar  K}^{0}\pi^{0}) $  which has 
been lowered  over the years. It was  
$(11.5\pm 1.0)\times 10^{-6} $ \cite{HFAG4} and exceeds 
the upper limit of $(10.8\pm 0.3)\times 10^{-6} $ in 2005, as shown in
 Eq.(\ref{BK0p0}) (the old $K\pi$ puzzle \cite{Chiang}). With our
value for ${\mathcal B}({\bar B}^{0}\to {\bar K}^{0}\pi^{0}) $ , the 
difference  would be reduced to $0.9\times 10^{-6}$
and becomes consistent with Eq.(\ref{sum}), within the present accuracy. We 
thus expect the experimental value for 
  ${\mathcal B}({\bar B}^{0}\to {\bar   K}^{0}\pi^{0}) $
would be lower than the current value and is close to our predicted value 
$9.0\pm 0.3\times 10^{-6} $. (The Belle value is
in fact $(9.2\pm 0.7^{+0.6}_{-0.7}) \times 10^{-6} $ \cite{HFAG}).
\bigskip
\section{Conclusion}
 By adding a mainly absorptive, penguin-like contribution to the 
$B\to K\pi$ decay QCDF amplitude, we show that QCDF could 
successfully predict the $B\to K\pi$ branching ratios. We obtain the 
correct magnitude
and sign for the ${\bar B}^{0}\to K^{-}\pi^{+} $ CP asymmetry. 
The $B^{-}\to K^{-}\pi^{0}$ CP asymmetry is predicted to be of the same sign
and  magnitude as that of ${\bar B}^{0}\to K^{-}\pi^{+} $  which
should be checked against more precise data. We also show that 
QCDF describes well the electroweak penguin contribution in the  
$B^{-}\to K^{-}\pi^{0}$ and ${\bar B}^{0}\to {\bar K}^{0}\pi^{0} $ 
branching ratios and show that the 
${\bar B}^{0}\to {\bar K}^{0}\pi^{0} $ measured branching ratio
should be  lower than the current value. 
\bigskip
\begin{center}
{\bf Acknowledgments} \end{center}
I would like to thank  Hsiang-nan Li, Hai-Yang Cheng,
Chuan-Hung Chen and the members of the Institute 
of Physics, Academia Sinica, Taiwan,  for the  warm hospitality 
and support for my stay at the Institute where part of the work
was done. Also thanks to P. Colangelo and  F. De Fazio for the  
warm hospitality and support at INFN, Bari, Italy. This work was 
supported in part by the EU contract No. MRTN-CT-2006-035482, "FLAVIAnet".


\end{document}